# Direct imaging of the energy transfer enhancement between two dipoles in a photonic cavity

Kaizad Rustomji,[1,2] Marc Dubois,[1] Boris Kuhlmey,[2] C. Martijn de Sterke,[2] Stefan Enoch,[1] Redha Abdeddaim,[1] Jérôme Wenger[1]

[1] Aix Marseille Univ, CNRS, Centrale Marseille, Institut Fresnel, Marseille, France

[2] Institute for Photonics and Optical Sciences (IPOS), School of Physics, University of Sydney NSW 2006, Australia

**Abstract**

Photonic cavities are gathering a large interest to enhance the energy transfer between two dipoles, with far-reaching consequences for applications in photovoltaics, lighting sources and molecular biosensing. However, experimental difficulties in controlling the dipoles' positions, orientations and spectra have limited the earlier work in the visible part of the spectrum, and have led to inconsistent results. Here, we directly map the energy transfer of microwaves between two dipoles inside a resonant half-wavelength cavity with ultrahigh control in space and frequency. Our approach extends Förster resonance energy transfer (FRET) theory to microwave frequencies, and bridges the gap between the descriptions of FRET using quantum electrodynamics and microwave engineering. Beyond the conceptual interest, we show how this approach can be used to optimize the design of photonic cavities to enhance dipole-dipole interactions and FRET.

## I. Introduction

Near-field energy transfer plays a key role in solar energy harvesting, [1–3] organic lighting sources [4–6] and molecular biosensing [7–9]. Förster resonance energy transfer (FRET) is the general designation to describe the nonradiative energy transfer from a donor dipole to an acceptor dipole [7,10,11]. Owing to its $1/R^6$ dependence with the donor-acceptor separation $R$, FRET has become a widely used biophysical technique to monitor the nanometer distance between two fluorescent labels and to investigate molecular interaction and conformation dynamics [9,12]. More recently, the energy transfer within resonant photonic cavities has been receiving increased interest, thanks to the ability to reach strong-coupling conditions, and to create new hybrid light-matter states [13–19].

The numerous applications of FRET and dipole-dipole interaction is a strong motivation to further enhance and control the energy transfer by tailoring the photonic environment [20–22]. Indeed, it has been established since the pioneering works of Purcell [23] and Drexhage [24] that the Local Density of Optical States (LDOS) controls the spontaneous emission of single dipolar emitters [25]. Engineering the LDOS using cavities [26,27], photonic crystals [28,29], plasmonic antennas [30,31] and metamaterials [32,33] has proven to be a versatile and powerful approach to enhance spontaneous emission. However, extending these results to the energy transfer between two separated dipoles remains a challenging and controversial issue [21,22]. The relationship between LDOS and FRET has been studied intensively, but the results lead to disparate and seemingly contradictory conclusions.

Resonant optical microcavities [20,34–36], nanoparticle arrays [37–39], single nanoparticles [40–43], subwavelength apertures [44,45] and nanoantennas [46–48] have been reported to enhance the energy transfer rate, while other studies on mirrors [49–52], microcavities [53,54], nanoparticles [55–57] and plasmonic antennas [58,59] reported no effect on the FRET rate.

A major source of the disparity in previous studies of FRET in photonic devices is the difficulty to control the donor and acceptor positions at levels far below the wavelength. This is extremely difficult in the optical regime, since it corresponds to sub-nanometer accuracy [48]. The same argument applies to the dipole orientations—though FRET is highly sensitive to the mutual orientation of the dipoles, most of the experiments involve a large degree of rotational flexibility for the fluorescent emitters [47]. Another problem stems from the large spectral bandwidth of most of the fluorescent emitters used for FRET, while theoretical work typically assumes single wavelength emission [49,53]. Lastly, optical measurements of FRET are based on indirect observations of acceptor fluorescence brightness, donor fluorescence quenching or donor lifetime reduction, but there is no direct measurement of the power transferred from the donor to the acceptor [60].

Here, we introduce a general methodology to analyze FRET at radio-frequencies using a microwave engineering approach. While the physics remains the same, performing experiments in the microwave domain allows us to measure the energy transfer between two dipoles directly with ultrahigh control of their position and orientation. We show here that the FRET rate and the two-point Green function $\overleftrightarrow{\boldsymbol{G}}(\boldsymbol{r}_D, \boldsymbol{r}_A)$ are proportional to the mutual impedance $Z_{21}$ of the two-port network describing the two dipoles. This result reconciles the different descriptions of dipole-dipole energy transfer using quantum electrodynamics, semi-classical electrodynamics and microwave engineering (Fig. 1a,b) and extends the analogy between these fields [61–63]. To demonstrate the relevance of this new method, we investigate the energy transfer inside a resonant cavity formed by two planar mirrors separated by a half wavelength. Similar photonic cavities have received a large interest owing to their potential to confine light and to enhance light-matter interactions in either the weak [20,34–36,53,54] or strong coupling regime [13–19]. Here, we directly map the positions and conditions leading to an enhancement of the dipole-dipole energy transfer inside the cavity. Our results explain the apparent contradictions found in earlier studies in the optical regime [20,54], and provide the first complete set of design guidelines for practical applications of FRET using cavities in photovoltaics and light sources [2–6]. The microwave engineering approach to FRET thus not only unifies different descriptions of the same phenomenon, it also provides a new practical tool to design and characterize energy transfer around photonic structures quantitatively. Moreover, it provides a direct experimental validation of the Green function theory.

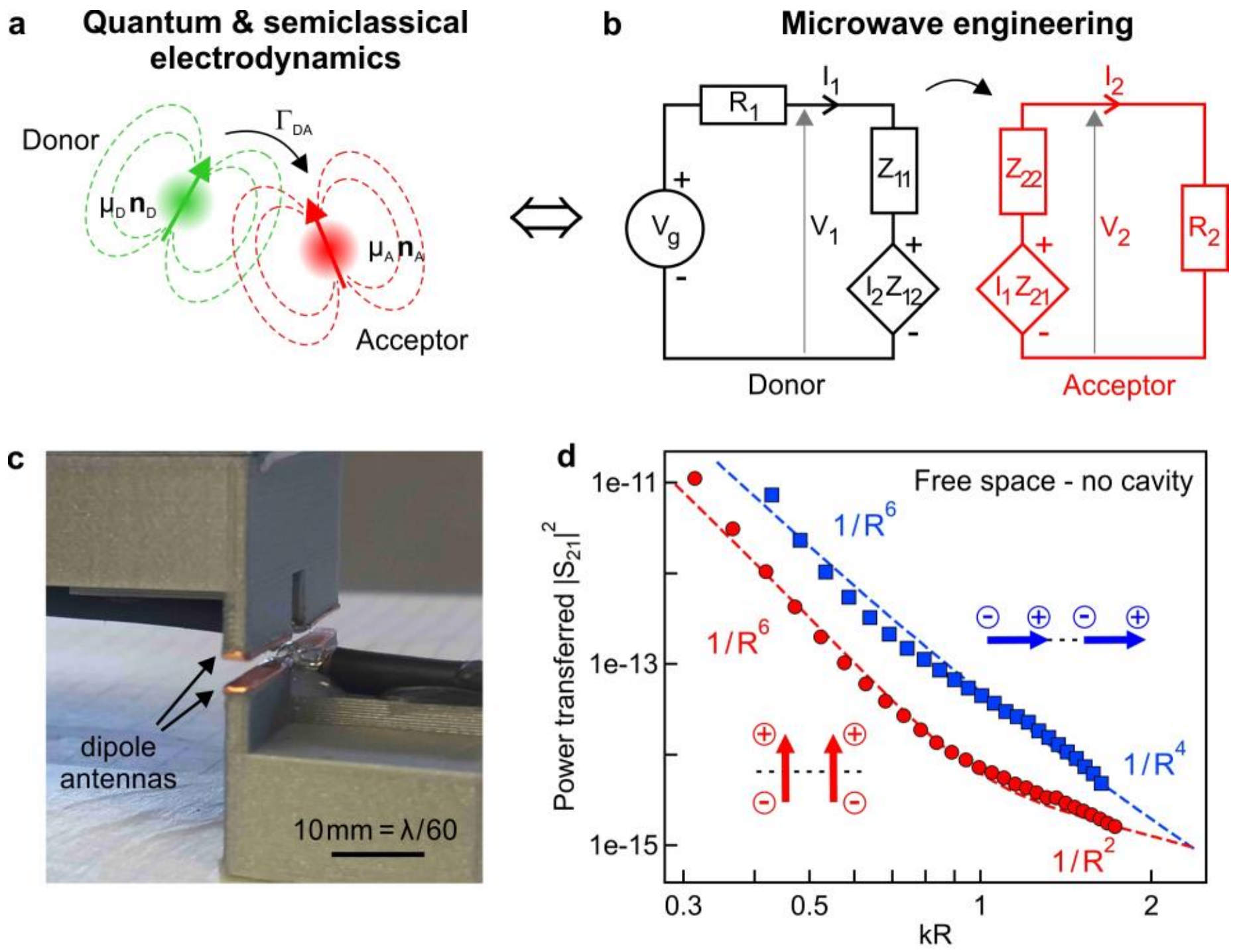

**Figure 1 (revised)**. (a) Quantum mechanical description, and (b) microwave engineering descriptions using a two-port network, of dipole-dipole energy transfer. (c) Photograph of the two dipoles set in parallel orientation. The dipole length is 30 mm and corresponds to λ/20. (d) Calibration of the power transferred versus the dipole-dipole separation R in free space. Markers are experimental data for parallel dipoles (red) and aligned dipoles (blue), dashed lines are theoretical predictions from Green's function theory [25].

## II. Theory

Within the semi-classical general description of dipole-dipole interaction, the power transferred by a donor D to the polarizable acceptor A can be written as [25]

$$P_{D\to A} = \frac{\omega}{2}\, Im\{\alpha_A\}\, |\, \boldsymbol{n}_A \cdot \boldsymbol{E}_D(\boldsymbol{r}_A)|^2, \tag{1}$$

where $\overleftrightarrow{\alpha_A} = \alpha_A\, \boldsymbol{n}_A \boldsymbol{n}_A$ is the acceptor's polarizability tensor and $\boldsymbol{E}_D(\boldsymbol{r}_A)$ is the electric field due to the donor dipole at the acceptor position. The donor field can be expressed using the Green function as

$$\boldsymbol{E}_D(\boldsymbol{r}_A) = \frac{\omega^2\, |\mu_D|}{c^2\, \varepsilon_0}\, \overleftrightarrow{\boldsymbol{G}}(\boldsymbol{r}_D, \boldsymbol{r}_A)\, \boldsymbol{n}_D\,, \tag{2}$$

where $|\mu_D|\, \boldsymbol{n}_D$ is the donor's dipole moment and $\overleftrightarrow{\boldsymbol{G}}(\boldsymbol{r}_D, \boldsymbol{r}_A)$ is the Green function at point $\boldsymbol{r}_A$ for a source located at $\boldsymbol{r}_D$. The power transferred is

$$P_{D\to A} = \frac{\omega^5\, |\mu_D|^2}{2\, c^4 \varepsilon_0^2}\, Im\{\alpha_A\} \left|\, \boldsymbol{n}_A \cdot \overleftrightarrow{\boldsymbol{G}}(\boldsymbol{r}_D, \boldsymbol{r}_A)\, \boldsymbol{n}_D \right|^2, \tag{3}$$

Note that the Green function is a two-point dyadic depending on the location of both the donor and the acceptor, and that derivations following quantum electrodynamics or semi-classical electrodynamics lead to the same result [22,25].

The Green function encompasses all the information about the photonic environment [25]. It can therefore be used to understand the origin of the discrepancy between the effect of the photonic environment on the LDOS and FRET rates. On one hand, the LDOS is proportional to the imaginary part of the Green function $Im\{\overleftrightarrow{\boldsymbol{G}}(\boldsymbol{r}_D, \boldsymbol{r}_D)\}$ at the emitter position: $LDOS = 6\,\omega/(\pi\, c^2)\, \boldsymbol{n}_D \cdot Im\{\overleftrightarrow{\boldsymbol{G}}(\boldsymbol{r}_D, \boldsymbol{r}_D)\} \cdot \boldsymbol{n}_D$, which depends only on the Green function at the source origin $\boldsymbol{r}_D$. This corresponds to the part of the radiation that is backscattered onto the source by the environment [25]. On the other hand, the FRET rate is proportional to the square modulus of the Green function of the donor when evaluated at the acceptor position $\left| \boldsymbol{n}_A \cdot \overleftrightarrow{\boldsymbol{G}}(\boldsymbol{r}_D, \boldsymbol{r}_A)\, \boldsymbol{n}_D \right|^2$, and corresponds to the donor power absorbed by the acceptor [22,25]. So while the Green function plays a role in determining both the LDOS and FRET, it does so in different manners. Therefore, no general relationship exists between the LDOS and energy transfer. This stresses the need for careful investigations of the influence of photonic environment on dipole-dipole energy transfer.

We now rephrase the dipole-dipole interaction by modeling it using a two-port network model (Fig. 1b) [64]. The donor corresponds to Port 1 with voltage $V_1$ and current $I_1$ driven by a source voltage $V_g$. It has a purely dissipative real-valued resistance $R_1$. The acceptor has no driving source and is purely passive; it is associated with Port 2 with voltage $V_2$, current $I_2$ and dissipative resistance $R_2$. The coupling between voltages and currents is summarized by the Z matrix as [64]

$$\begin{pmatrix} V_1 \\ V_2 \end{pmatrix} = \begin{pmatrix} Z_{11} & Z_{12} \\ Z_{21} & Z_{22} \end{pmatrix} \begin{pmatrix} I_1 \\ I_2 \end{pmatrix}. \tag{4}$$

The full Z matrix can be measured at microwave frequencies using a vector network analyzer [63]. We now show how it relates to the power transferred between dipoles and to the Green function $\overleftrightarrow{\boldsymbol{G}}(\boldsymbol{r}_D, \boldsymbol{r}_A)$.

With this definition of the Z matrix, the voltage induced in the acceptor circuit by the donor is $I_1\, Z_{21}$ (Fig. 1b). This induced source dissipates a certain power which corresponds to the power transferred to the acceptor:

$$P_{1\to 2} = \frac{1}{2} Re\{\, (I_1\, Z_{21})\, I_2^*\}. \tag{5}$$

Kirchhoff's voltage law in the acceptor circuit allows to write the link between the two currents as $I_2 = -I_1\, Z_{21}/\,(R_2 + Z_{22})$ . Therefore, the power transferred to the acceptor is

$$P_{1\to 2} = \frac{1}{2} |I_1|^2\, |Z_{21}|^2\, \frac{Re\{R_2 + Z_{22}\}}{|R_2 + Z_{22}|^2}. \tag{6}$$

This expression shows that the key element bridging the donor and acceptor is the mutual impedance $Z_{21}$.This impedance is defined as the ratio of voltage $V_{2o}$ induced in port 2 by the current $I_1$ in the absence of a current $I_2$ [64]

$$Z_{21} = \left. \frac{V_{2o}}{I_1} \right|_{I_2=0}. \tag{7}$$

As the acceptor antenna length $l_A$ is much shorter than the wavelength, the electric field $\boldsymbol{E}_D(\boldsymbol{r}_A)$ generated by the donor can be taken to be uniform over the acceptor antenna. Therefore the voltage $V_{2o}$ scales linearly with the donor's electric field: $V_{2o} = \boldsymbol{n}_A \cdot \boldsymbol{E}_D(\boldsymbol{r}_A)\, l_A$. Likewise, the current $I_1$ is related to the donor's dipole moment by $I_1 = -\, i\, \omega\, \mu_D / l_D$ , where $l_D$ is the donor antenna length. [65] Thus the impedance $Z_{21}$ can be expressed as

$$Z_{21} = \frac{\boldsymbol{n}_A \cdot \boldsymbol{E}_D(\boldsymbol{r}_A)\, l_A\, l_D}{-\, i\, \omega\, \mu_D} = \frac{i\, \omega\, l_A\, l_D}{c^2\, \varepsilon_0}\; \boldsymbol{n}_A \cdot \overleftrightarrow{\boldsymbol{G}}(\boldsymbol{r}_D, \boldsymbol{r}_A)\, \boldsymbol{n}_D. \tag{8}$$

This equation is our main result: the Green function $\overleftrightarrow{\boldsymbol{G}}(\boldsymbol{r}_D, \boldsymbol{r}_A)$ governing the energy transfer is proportional to the mutual impedance $Z_{21}$ of the two-port network. While previous works on the analogy between microwave engineering and quantum electrodynamics focused on the LDOS and the Purcell factor [61–63], our Eq. (8) extends the analogy to the important case of dipole-dipole interaction. As we detail in the Supplemental Information Table S1, the link between the semi-classical formalism (Eq. 3) and the two-port network (Eq. 6) can be taken further to connect all physical quantities defining the dipole-dipole interaction with equivalent parameters of two-port networks.

Having established the relationship between $Z_{21}$ and $\overleftrightarrow{\boldsymbol{G}}(\boldsymbol{r}_D, \boldsymbol{r}_A)$, we can now express the *energy transfer enhancement*, quantifying the influence of the photonic environment on the energy transfer as [22]

$$F_{ET} = \frac{P_{D\to A}}{P^0_{D\to A}} = \frac{\left| \boldsymbol{n}_A \cdot \overleftrightarrow{\boldsymbol{G}}(\boldsymbol{r}_D, \boldsymbol{r}_A)\, \boldsymbol{n}_D \right|^2}{\left| \boldsymbol{n}_A \cdot \overleftrightarrow{\boldsymbol{G}^0}(\boldsymbol{r}_D, \boldsymbol{r}_A)\, \boldsymbol{n}_D \right|^2} = \frac{|Z_{21}|^2}{|Z^0_{21}|^2}, \quad (9)$$

where the subscript 0 in the denominator denotes the a homogeneous environment. The relative change of the two-point Green function connecting two dipoles can thus be obtained by complex mutual impedance $Z_{21}$ measurements using a vector network analyzer. Alternatively, the scattering coefficient $S_{21}$ can be used instead of the impedance $Z_{21}$ as both Z- and S- parameters are related to each other [66]. The scattering coefficient $S_{21}$ gives the power dissipated in the resistance $R_2$

$$|S_{21}|^2 = \frac{1}{2}\, |I_2|^2 R_2 = \frac{1}{2} |I_1|^2\, |Z_{21}|^2\, \frac{R_2}{|R_2 + Z_{22}|^2}. \quad (10)$$

As the antenna length is short compared to the wavelength, $Re\{Z_{22}\} \ll R_2$ [62], $|S_{21}|^2$ is approximatively equal to the power transferred to the acceptor $P_{1\to 2}$ (Eq. 6). Therefore $S_{21}$ or $Z_{21}$ can be equivalently used to determine the energy transfer enhancement $F_{ET}$.

## III. Results and discussion

*Free space results*

We start by validating our approach by measuring the power transfer in free space as function of the dipole separation. The free space Green function is completely known and analytical formulas can be derived for the transferred power $P_{D\to A}$ [25]. This makes the free space configuration a relevant test for our approach. For the experiments, the two dipoles (Fig. 1c) are connected to a vector network analyzer which measures the amplitude and phase of the mutual impedances $Z_{21}$ and $S_{21}$. The power transferred from the donor to the acceptor $|S_{21}|^2$ can thus be recorded as the distance $R$ between dipoles varies.

Our experimental results recover all the features expected for dipole-dipole energy transfer in free space (Fig. 1d). In the near-field, $kR = 2\pi R/\lambda < 1$, the transfer rate scales as $1/R^6$ for the both dipole orientations. The far-field behavior ($kR > 1$) depends on the relative orientation between dipoles. It falls as $1/R^2$ for parallel dipoles and $1/R^4$ for aligned dipoles. We stress that the lines in Fig. 1d stem from the analytical expression of the Green function [25] and are not numerical fits. The only free parameter is a scaling factor for the amplitude, and the same factor was used for both dipole orientations. Note the slight deviations from the theory when the dipoles are aligned. We believe this is because the aligned configuration is more sensitive to the finite size and shape of the dipoles, especially at their tips. Hereafter, we focus on the parallel configuration, which appears to be less sensitive to this effect.

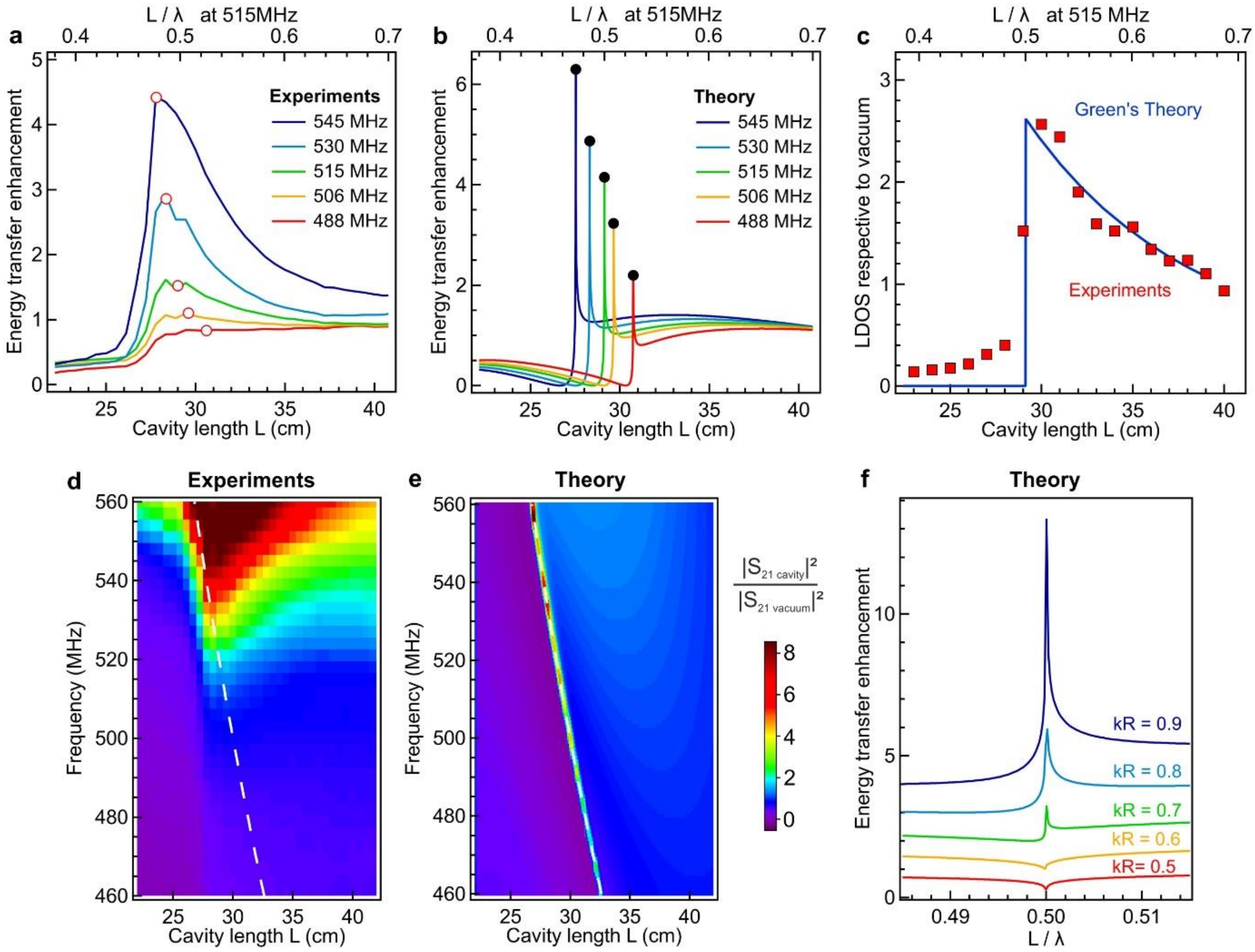

**Figure 2**. Dipole-dipole energy transfer in a cavity. (a) Measured energy transfer enhancement factor (Eq. 9) between two dipoles versus cavity length, for different frequencies. The two dipoles are oriented parallel to each other and parallel to the mirror plates forming the cavity. From top to bottom, the different frequencies correspond to kR values of 0.87, 0.84, 0.82, 0.80 and 0.77 respectively. The white disks indicate the cavity length that maximizes the energy transfer. It shifts to lower values with increasing frequency. (b) Simulations using Eq. (9) of the experimental results in (a). Due to the finite size effects, the sharp peaks obtained at resonance (black disks) are smoothed in the experimental data in (a), but the main features (peak position and relative amplitude) are preserved as the frequency is tuned. (c) LDOS enhancement (Purcell factor) versus cavity length at 515 MHz, for a single dipole at 11.2 cm from the nearest mirror with orientation parallel to the mirrors. The blue line is the prediction from Green's function theory while the red squares are experimental data. (d,e) Energy transfer enhancement versus cavity length and frequency. The experimental data (d) and the simulations (e) share the same color scale. The white dashed line represents the expected trend $L = \lambda/2$ for the resonance condition. (f) Simulated evolution of the energy transfer enhancement around the resonance for increasing dipole-dipole separations R as function of the normalized cavity length $L/\lambda$. For clarity, the curves are vertically shifted by +1.

*Cavity results*

Having validated the approach, we now set the dipoles into a cavity formed by two metallic plate mirrors. The cavity length $L$ is controlled and scanned around the first resonance condition $L = \lambda/2$ corresponding to a half-wavelength cavity. The donor dipole is fixed at 11.2 cm from the nearest mirror, and the dipole-dipole separation is maintained at 7.6 cm, corresponding to the near-field regime ($kR < 1$) for all frequencies used. The emission frequency $f$ is also scanned so as to change the parameter $kR = 2\pi R/\lambda = 2\pi Rf/c$ without touching the geometrical configuration of the dipoles. Figure 2 summarizes our main results for the energy transfer enhancement $F_{ET}$ in a resonant cavity. A clear enhancement factor up to $F_{ET} = 4.4 \times$ for 545 MHz frequency is observed near the resonance condition $L = \lambda/2$ (Fig. 2a). For the short cavity length below the resonance $L < \lambda/2$, the energy transfer is suppressed $F_{ET} < 1$. For larger cavities above the resonance $L > \lambda/2$, the energy transfer is moderately increased as $F_{ET}$ is slightly greater than 1. The maximum enhancement is always seen around the resonance condition (white dots in Fig. 2a) which shifts towards shorter cavity lengths as the frequency is increased. All these features are consistent with simulations following Green's theory (Fig. 2b), though the finite size of the dipole probes and the finite size of the cavity tend to smooth the experimental data as compared to the theoretical predictions. Note that the peak positions coincide with the resonance condition $L = \lambda/2$ for both the experimental and the simulated data (Supplemental Material Fig. S1).

Theoretically, the energy transfer enhancement follows an asymmetric Fano-like line-shape as the cavity length is varied (Fig. 2b) [67]. This behavior is typical of the interference between the contributions from a continuum of states (the energy directly transferred from the donor to the acceptor as in free space) and a discrete resonant state (the contribution from the cavity). Near the resonance, the donor field backscattered by the cavity rapidly changes phase, creating the asymmetric Fano line-shape. The experimental observations follow the same trend (Fig. 2a), although the resonance peak appears less sharp due to the finite size of the dipole antennas and the cavity.

It is interesting to relate the energy transfer enhancement $F_{ET}$ to the LDOS enhancement or Purcell factor (Fig. 2c). For a dipolar emitter in the cavity center with parallel orientation, the Green function theory predicts a maximum LDOS enhancement up to $3\times$ at the resonance condition $L = \lambda/2$. While both the energy transfer and the LDOS share the same resonance condition, there are also marked differences between them. Below resonance, there is a cut-off for the dipole radiation and the LDOS vanishes. However, the energy transfer remains nonzero, though it is suppressed. The limiting case for short cavities is especially interesting: theoretically, the LDOS is zero but the energy transfer enhancement tends to $F_{ET} = 1$, as if there was no cavity at all (Fig. 2b).

The maximum energy transfer enhancement increases with increasing frequency (Fig. 2a,b). To better represent this phenomenon, we show in Fig. 2d and 2e 2D maps of the energy transfer enhancement versus the emission frequency and the cavity length. We again find consistency between the experimental (Fig. 2d) and theoretical results (Fig. 2e), within the limitations due to the finite antenna size. The trend observed when the frequency is tuned suggests a dependence of the parameter $kR$. As the distance $R$ between the dipoles is fixed, tuning the frequency $f$ is equivalent to changing the parameter $kR = 2\pi Rf/c$. We therefore perform simulations for different $kR$ values (Fig. 2f). For small dipole-dipole separations ($kR < 0.6$), the energy transfer is always near its value in free space and can even be suppressed near the resonance. However for larger distances, a peak starts to appear at the resonance condition $L = \lambda/2$ [34]. The larger the dipole separation, the larger the influence of the photonic environment. This is consistent with earlier optical FRET studies using nanostructures [44–46].

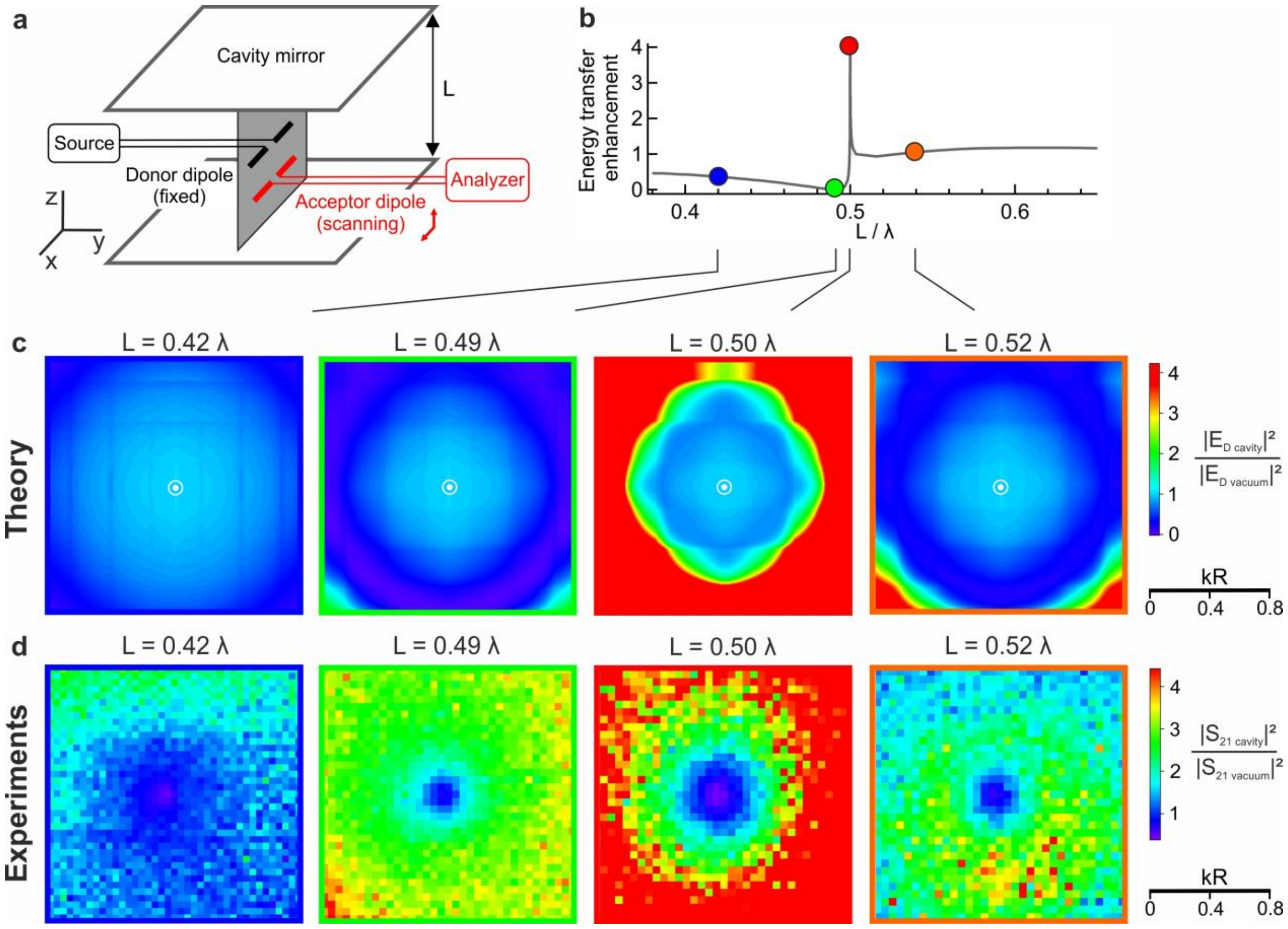

**Figure 3 (revised).** Spatial maps of the energy transfer enhancement inside the cavity. (a) Sketch of the experiment. The donor dipole (black) is fixed at 11.2 cm from the nearest mirror, while the acceptor dipole (red) is scanned to record the energy transfer at a fixed frequency. All data are then normalized to the same experiment performed without the cavity. (b) Simulation of the energy transfer enhancement versus cavity lengths, with the lengths used in the experiments indicated by colored dots. For this simulation, a dipole-dipole distance of 7.6 cm (corresponding to $kR$ = 0.8) was used. (c) Simulated maps of the energy transfer enhancement inside the cavity for two dipoles oriented along X and for different cavity lengths below, at, and above resonance. The white dot indicates the donor dipole position. Maps computed for a larger spatial domain up to kR=2 are shown in the Supplemental Material Fig. S2. (d) Measured maps of the energy transfer enhancement for cavity lengths below, at, and above the resonance condition. The central dip (purple spot) corresponds to the donor dipole position.

*Mapping of the energy transfer rate*

We now directly map the energy transfer versus acceptor position inside the cavity. The cavity length is also controlled so as to allow us to work below, at or above the resonance. This is the first time that energy transfer is directly probed as function of the donor-acceptor separation inside a photonic system. The positioning accuracy is remarkably high: every pixel in the measured images corresponds to a distance of λ/130 at the operating frequency of 500 MHz, corresponding to a wavelength of 60 cm. Figure 3 summarizes our main results. The maximum enhancement is again found when the cavity length meets the resonance condition $L = \lambda/2$. As inferred from Fig. 2, we directly see that increasing the donor-acceptor separation $R$ tends to increase the enhancement factor. Energy transfer enhancement factors $F_{ET}$ above $4\times$ are clearly seen on both the simulated and experimental data for the distance range corresponding to $kR > 0.7$. Numerical simulations predict even higher values above one hundred fold, though for separations exceeding $kR > 1$ where radiative (far-field) coupling dominates the energy transfer process (Supplemental Material Fig. S2 and S3). In the very near-field of the donor, for separations $kR$ below 0.2, a blue spot is found around the donor position for all the cavity lengths. This corresponds to conditions where the energy transfer is close to the free space result and the cavity brings no improvement.

For short cavity lengths below the cut-off $L < \lambda/2$, the energy transfer enhancement is close to unity and is uniformly distributed inside the cavity. Although the LDOS is heavily suppressed under these conditions (Fig. 2c), the energy transfer is not, and we can recover a distribution similar to the free space result. For cavity lengths near the resonance, the donor energy has a more complex distribution which can lead to transfer rates exceeding that in homogeneous space. This configuration appears especially interesting for the FRET applications dealing with light harvesting and light-emitting sources where donors and acceptors are broadly distributed over the cavity volume. Further images for different configurations and larger length scales are shown in the Supplemental Material Fig. S2 and S3.

*Discussion*

To help with the design of photonic cavities to enhance dipole-dipole energy transfer, we present in Fig. 4a,b two-dimensional graphs showing the energy transfer enhancement $F_{ET}$ as functions of the normalized cavity length $L/\lambda$ and donor-acceptor separation $kR = 2\pi R/\lambda$. The maps also indicate the boundaries where the LDOS is enhanced or quenched (Purcell factor = 1) and where the energy transfer is enhanced or quenched ($F_{ET} = 1$). At least three main conclusions can be drawn from this figure as we discuss below.

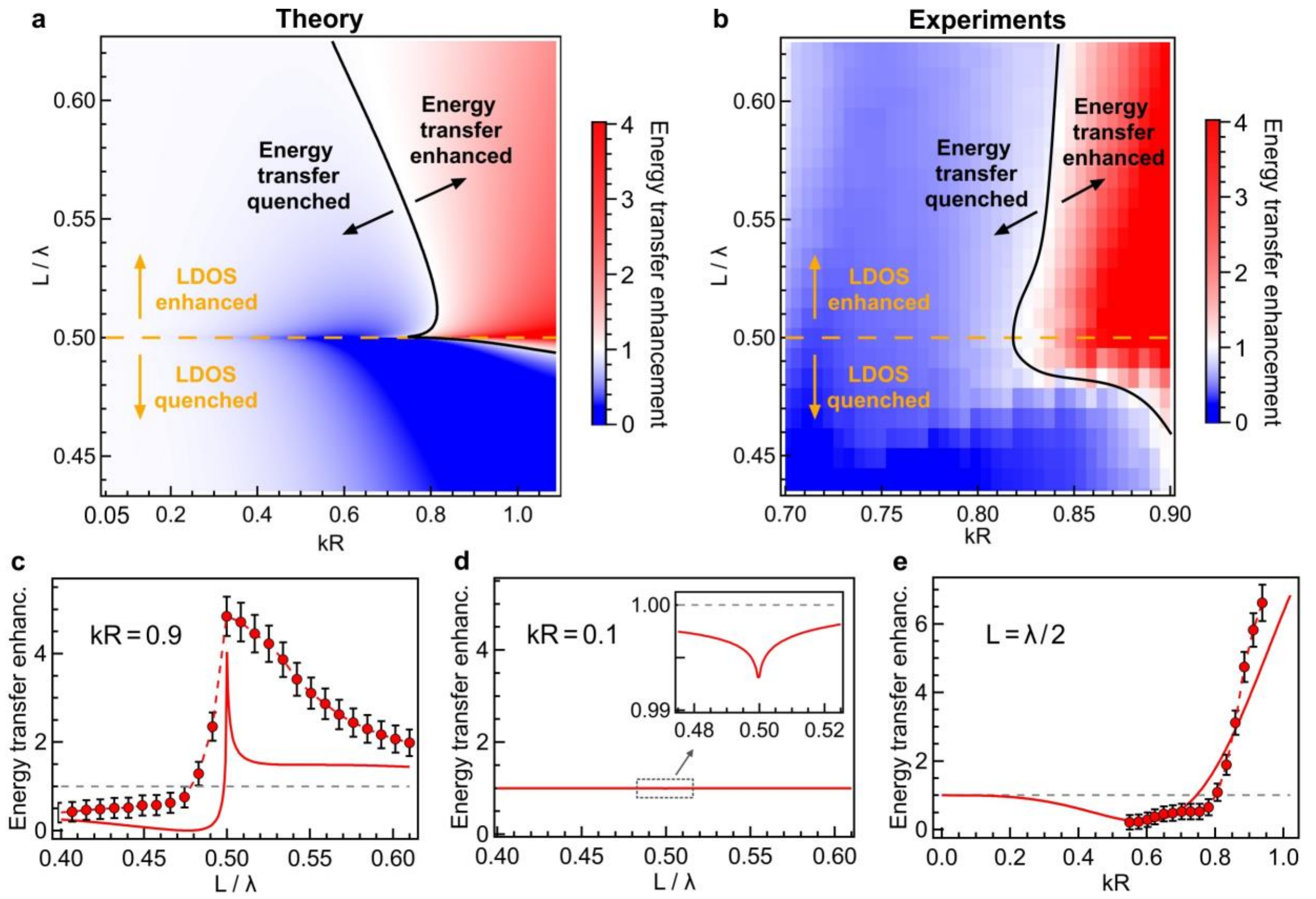

**Figure 4 (revised).** Two-dimensional maps summarizing the dipole-dipole energy transfer enhancement $F_{ET}$ as functions of the normalized cavity length $L/\lambda$ and dipole-dipole separation $kR$. (a) Simulated map and (b) experimental map for the dipoles in parallel orientation (as in Fig. 3a). The donor dipole is fixed at 11.2 cm from the nearest mirror. In (a,b) the black lines show the contour where the energy transfer enhancement equals unity. The yellow dashed lines show the boundary between LDOS enhancement and quenching at $L/\lambda = 0.5$. The theoretical and experimental maps share the same colorscale. (c) Energy transfer enhancement factor obtained for kR = 0.9 as function of the normalized cavity length L/λ. The solid dashed line represents the numerical prediction while the dots are experimental data. The grey dashed line indicates the level where $F_{ET} = 1$. (d) Same as (c) for kR = 0.1. The insert is a close-up view of the region enclosed in the dashed rectangle. (e) Energy transfer enhancement factor at the cavity resonant length L=λ/2 for increasing dipole-dipole separation kR. In (c,e) the experimental error bars are estimated by summing a 5% experimental measurement error plus a 0.2 fixed contribution to represent the error in estimating the reference level in free space.

First, the LDOS enhancement and energy transfer enhancement follow clearly different contours. There exist regions where (i) both the LDOS and energy transfer are quenched; (ii) the LDOS is enhanced but the energy transfer is quenched; (iii) the LDOS is quenched while the energy transfer is enhanced; and (iv) both the LDOS and the energy transfer are enhanced. Even for the conceptually simple case of a cavity formed by two planar mirrors, Fig. 4a shows that there is no general relationship between FRET and LDOS and that under some conditions both FRET and LDOS may seem correlated while under different conditions they may appear totally independent [50].

Secondly, for short dipole-dipole separations $kR < 0.3$, the energy transfer is always close to the free space result and the cavity has a negligible effect on the energy transfer rate, irrespective of the cavity length and LDOS (Fig. 4d,e). This explains the experimental observations for a single FRET pair in an optical microcavity. [54] Moreover, the fact that the energy transfer is only slightly changed by the cavity when the length is below the resonance $L < \lambda/2$ and the donor-acceptor separation is short $kR < 0.2$ has a very interesting consequence. In this configuration, spontaneous emission is prohibited and the LDOS vanishes so the donor can only transfer its energy to the acceptor. The energy transfer efficiency in this case thus approaches 100% for a broad range of donor-acceptor separations (Supplemental Material Fig. S4). This suggests new routes to dramatically improve the FRET efficiency, as we discuss in the section 6 of the Supplemental Material. Similarly, photonic crystals were proposed to suppress the LDOS and enhance the dipole-dipole energy transfer [49]. Our results for planar mirror cavities have the additional advantages of easier experimental implementation and broader spectral range, yet they require the donor dipole to have an orientation parallel to the mirrors in order to suppress the LDOS.

Thirdly, the maximum energy transfer enhancement is reached for large donor-acceptor separations $kR > 0.8$. Under this condition and at the resonance cavity length, the enhancement factor can theoretically be very large, exceeding two orders of magnitude (Fig. 4c,e). This affects experiments using spatially large layers of emitters [20,35,36], where the donor can couple efficiently not only with the nearest acceptor, but also with a large number of more distant acceptors. Consequently, the data averaged over large ensemble of emitters [20,35,36] can significantly differ from the single molecule results [54]. Additionally, collective effects may also play a non-negligible role in FRET [68]. This phenomenon appears also very promising to couple efficiently molecular emitters and antennas over long distances exceeding several wavelengths [69,70].

## IV. Conclusion

We have developed the concept of mutual impedance of a two-port network and related it to the quantum and semiclassical electrodynamics model of dipole-dipole energy transfer. This unified description of energy transfer allows to analyze microwave experiments and connect the results with the well-known Förster's formalism and Green's function theory. It simultaneously provides a deeper understanding of energy transfer and acts as a practical tool to design and characterize photonic devices with enhanced dipole-dipole interaction. Additionally, it provides a direct experimental validation of Green's theory.

A major advantage of our approach is that it allows the direct measurement of the power transferred from the donor to the acceptor with excellent precision and control of the dipoles positions, orientations and emission spectra. To demonstrate the effectiveness of our approach, we explored the energy transfer inside a resonant cavity and derived the conditions where the dipole-dipole energy transfer is enhanced. Performing experiments using microwaves allows us to investigate a much wider set of conditions than in the optical regime. Our general results explain why earlier works in the optical regime led to apparently contradictory conclusions, and also provides new practical guidelines to improve FRET applications with resonant cavities. Altogether, this work preludes a new class of studies investigating FRET inside inhomogeneous environments at ultrahigh spatial resolution.

**Acknowledgments**

The authors thank John Sipe for helpful discussions. This research was conducted within the context of the International Associated Laboratory "ALPhFA: Associated Laboratory for Photonics between France and Australia." This work has received funding from the European Union's Horizon 2020 Research and Innovation programme under Grant Agreement No 736937, from the Agence Nationale de la Recherche (ANR) under grant agreement ANR-17-CE09-0026-01, and from Excellence Initiative of Aix-Marseille University - A*MIDEX, a French "Investissements d'Avenir" programme.

**Appendix A: Experimental setup**

Dipole supports are designed with CAD software and 3D printed in polylactic acid polymer (PLA) with an Ultimaker 2 printer. Copper tape is deposited to obtain the two branches of the dipole then soldered to a coaxial cable. The dipole dimensions are 30 mm length, 2 mm width, 0.1 mm thickness, with a gap between branches of 3 mm. For the energy transfer measurements, the cavity is formed by two planar mirrors made of 33 µm thick copper deposited on dielectric plaques. Each mirror has dimensions of 180 x 120 cm² corresponding to 3λ x 2λ. For the LDOS measurement, we further refine the cavity to use two bigger iron plaques of 1 mm thickness and 300 x 200 cm² (5λ x 3.3λ). All the mirrors have been stiffened by braces and can be considered as perfect electric conductors (PEC) at the microwave frequencies used here. The parallelism between the two mirrors is controlled by the cavity design to better than 1°. For all the experiments reported here, the donor dipole is located in the center of the mirror at 11.2 cm distance from the nearest mirror. This configuration is used to break the symmetry along the Z axis and better reveal the spatial maps inside the cavity (Supplemental Material Fig. S2) but it does not significantly change our conclusions as compared to the configuration where the donor dipole is located exactly in the cavity center. Impedance and S-parameters are obtained with a vector network analyzer (model Anritsu MS2036C) which has been fully calibrated over the frequency bandwidth. Results are obtained with an IF filter of 10 Hz (acquisition rate) and averaged 10 times. LDOS measurement are obtained by increasing the distance between the two metallic plates composing the cavity. The dipole position within the cavity is set with polystyrene foam spacers. The FRET maps are obtained by scanning the acceptor dipole position with a multi axis motorized translation stage with a 5x5 mm² spatial sampling.

**Appendix B: Numerical simulations**

Numerical simulations for S-Parameters, as well as field maps, were performed using commercial finite difference time domain (FDTD) solver, CST Microwave studio (2015). Each dipole antenna was modeled as two perfect electric conducting (PEC) cylinders, with a small gap between the two cylinders. The dipoles were excited using "discrete ports", as available in CST, connecting the gap. Since the antenna lengths are much smaller than the wavelength, the mesh was refined near the dipole antennas. To truncate the computational domain perfectly matched layers (PMLs) were applied on all sides. For simulations inside the cavity, the cavity plates were modeled as two perfectly electric conducting sheets of thickness 0.1 mm with each plate of edge length $L>2\lambda$, ensuring the finite sized effects could be safely ignored.

# Supplemental material for

# Direct imaging of the energy transfer enhancement between two dipoles in a photonic cavity

Kaizad Rustomji,[1,2] Marc Dubois,[1] Boris Kuhlmey,[2] C. Martijn de Sterke,[2]

Stefan Enoch,[1] Redha Abdeddaim,[1] Jérôme Wenger[1]

[1] Aix Marseille Univ, CNRS, Centrale Marseille, Institut Fresnel, Marseille, France

[2] Institute for Photonics and Optical Sciences (IPOS), School of Physics, University of Sydney NSW 2006, Australia

**Contents:**

## 1. Relationships between two-port network model and classical dipole-dipole electrodynamics

In this section, we link the parameters used in the two-port network model to the physical quantities used to describe the dipole-dipole interaction. The key assumption is that the antenna lengths $l_D$ $l_A$ for the donor and acceptor are both small compared to the wavelength (point source approximation).

Within the classical electrodynamics formalism, the power transferred by the donor to the polarizable acceptor is [1]

$$P_{D\to A} = \frac{\omega^5\, |\mu_D|^2}{2\, c^4 \varepsilon_0^2}\, Im\{\alpha_A\} \left| \boldsymbol{n}_A \cdot \overleftrightarrow{\boldsymbol{G}}(\boldsymbol{r}_D, \boldsymbol{r}_A)\, \boldsymbol{n}_D \right|^2 \qquad \text{(S1)}$$

where $|\mu_D|\, \boldsymbol{n}_D$ is the donor's dipole moment, $\alpha_A\, \boldsymbol{n}_A \boldsymbol{n}_A$ the acceptor's polarizability tensor and $\overleftrightarrow{\boldsymbol{G}}(\boldsymbol{r}_D, \boldsymbol{r}_A)$ the Green function at point $\boldsymbol{r}_A$ for a source located at $\boldsymbol{r}_D$.

With the two-port network model, the power transferred from the donor to the acceptor is

$$P_{1\to 2} = \frac{1}{2} |I_1|^2\, |Z_{21}|^2\, \frac{Re\{R_2 + Z_{22}\}}{|R_2 + Z_{22}|^2} \cong |S_{21}|^2 \qquad \text{(S2)}$$

To link the powers given by Eq. (S1) and (S2), we must relate each term in the equations. First, the current $I_1$ is related to the donor's dipole moment $\mu_D$ [2–4]:

$$\mu_D = -\frac{I_1\, l_D}{i\, \omega} \qquad \text{(S3)}$$

The acceptor polarizability $\alpha_A$ takes the form of an electrically small lossy scatterer [2,5]

$$\frac{1}{\alpha_A} = -\frac{i\, \omega}{{l_A}^2} (Z_{22} + R_2) \qquad \text{(S4)}$$

So that we can rewrite

$$Im\{\alpha_A\} = -\frac{{l_A}^2}{\omega} \frac{Re\{Z_{22} + R_2\}}{|Z_{22} + R_2|^2} \qquad \text{(S5)}$$

We have shown in the main text that the formal definition of the impedance $Z_{21} = \left.\frac{V_{2o}}{I_1}\right|_{I_2=0}$ relates to the Green function:

$$Z_{21} = \frac{i\, \omega\, l_A\, l_D}{c^2\, \varepsilon_0}\, \boldsymbol{n}_A \cdot \overleftrightarrow{\boldsymbol{G}}(\boldsymbol{r}_D, \boldsymbol{r}_A)\, \boldsymbol{n}_D \qquad \text{(S6)}$$

By combining Eqs. (S3), (S5) and (S6), we retrieve a full equivalence between the powers given by either formalism Eq. (S1) or (S2). Table S1 summarizes our main equations.

**Table S1.** Equivalence between classical dipole-dipole electrodynamics and two-port network

| | Classical electrodynamics | Two-port network |
|---|---|---|
| **Power transferred** | $P_{D\rightarrow A} = \frac{\omega^5 \lvert\mu_D\rvert^2}{2\, c^4 \varepsilon_0^2} Im\{\alpha_A\} \left\lvert \boldsymbol{n}_A \cdot \overleftrightarrow{\boldsymbol{G}}(\boldsymbol{r}_D, \boldsymbol{r}_A)\, \boldsymbol{n}_D \right\rvert^2$ | $P_{1\rightarrow 2} = \frac{1}{2} \lvert I_1\rvert^2 \lvert Z_{21}\rvert^2 \frac{Re\{R_2 + Z_{22}\}}{\lvert R_2 + Z_{22}\rvert^2} \cong \lvert S_{21}\rvert^2$ |
| **Donor dipole moment** | $\mu_D$ | $-\frac{I_1\, l_D}{i\, \omega}$ |
| **Acceptor polarisability** | $Im\{\alpha_A\}$ | $-\frac{{l_A}^2}{\omega} \frac{Re\{Z_{22} + R_2\}}{\lvert Z_{22} + R_2\rvert^2}$ |
| **Green's function** | $\boldsymbol{n}_A \cdot \overleftrightarrow{\boldsymbol{G}}(\boldsymbol{r}_D, \boldsymbol{r}_A)\, \boldsymbol{n}_D$ | $\frac{Z_{21}}{l_A\, l_D} \frac{c^2\, \varepsilon_0}{i\, \omega}$ |

## 2. Concurrence between experimental and theoretical resonance frequencies

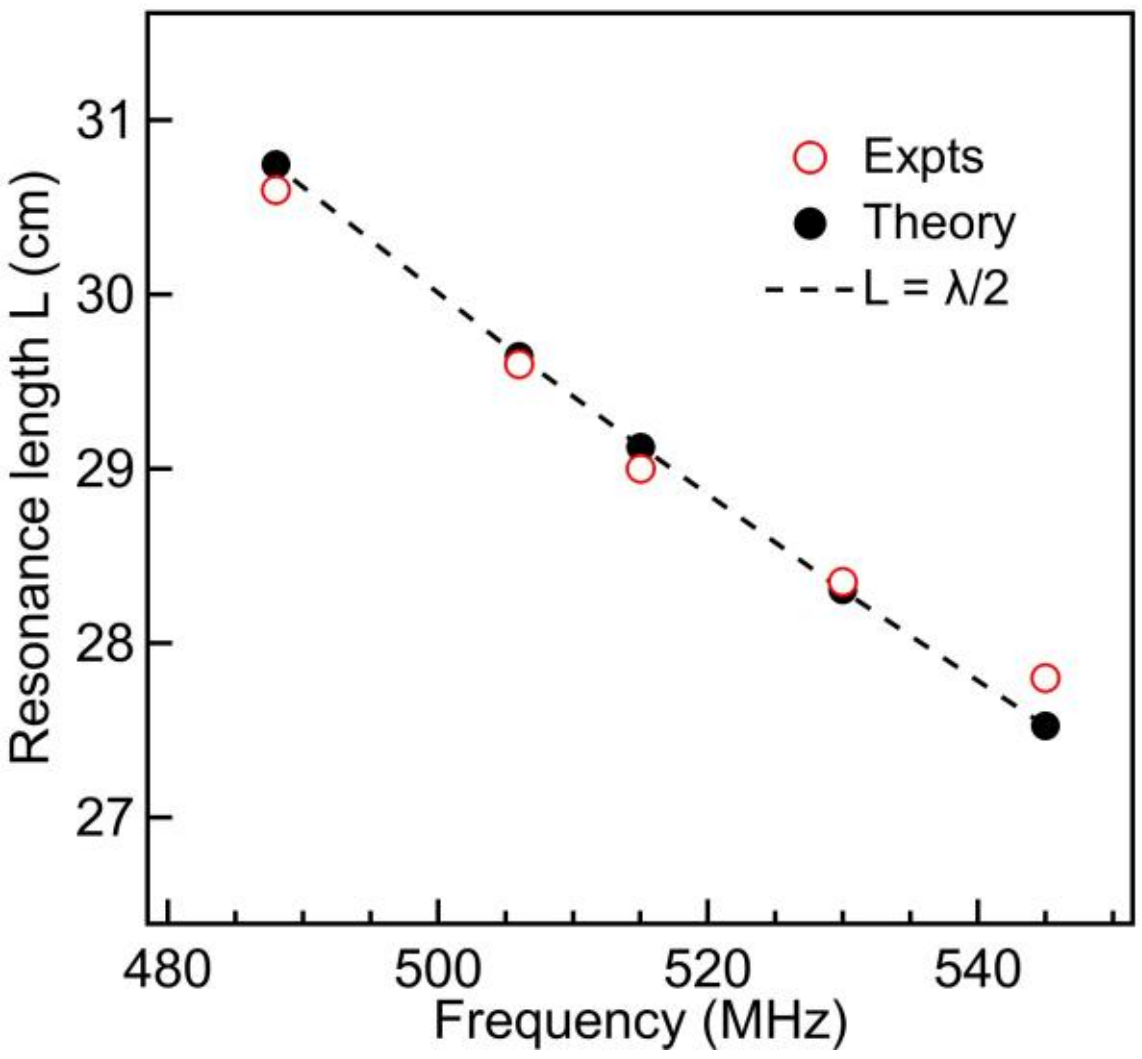

**Figure S1.** Evolution of the resonance length for the cavity as a function of the emission frequency, measured from the graphs in Fig 2a,b. Both the experimental data (white markers) and the simulated data (black markers) agree well with the expected values L = $\lambda/2$ (dashed line).

### 3. Simulated maps of the dipole-dipole energy transfer enhancement inside the cavity

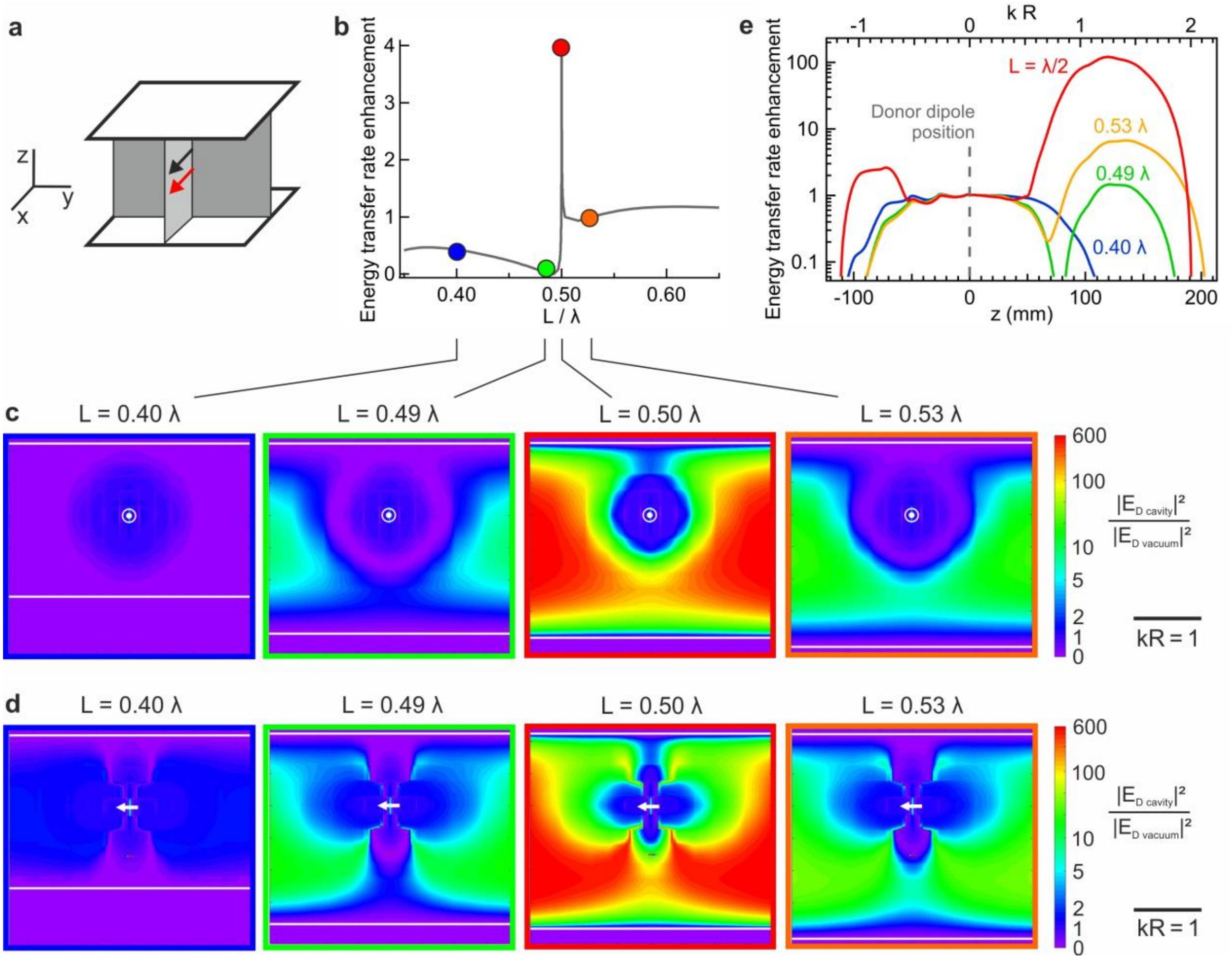

**Figure S2.** Simulated maps of the dipole-dipole energy transfer enhancement inside the cavity. (a) Sketch of the simulations, the donor dipole (black) is fixed at a 11.2 cm distance of the nearest mirror, while the acceptor dipole (red) is moved in the YZ or XZ plane to compute the maps (frequency 500 MHz, wavelength 60 cm). (b) Selection of the different cavity lengths (color dots) on the graph of energy transfer enhancement as function of the normalized cavity length. For this simulation, a dipole-dipole distance of 7.6 cm (corresponding to $kR$ = 0.8) was used. (c) YZ maps of the dipole-dipole energy transfer enhancement inside the cavity for two dipoles oriented along X and for different cavity lengths below, at, and above resonance. The white dot indicates the donor dipole position. Note that the domain size shown here is about twice larger as the one shown in Fig. 3c. (d) Same as (c) for the XZ plane. (e) Cross-cut view along the vertical (Z) direction of the energy transfer enhancement for different cavity lengths and different z positions of the acceptor dipole. Note the logarithmic vertical scale.

4. **Long range enhancement of energy transfer inside a photonic cavity**

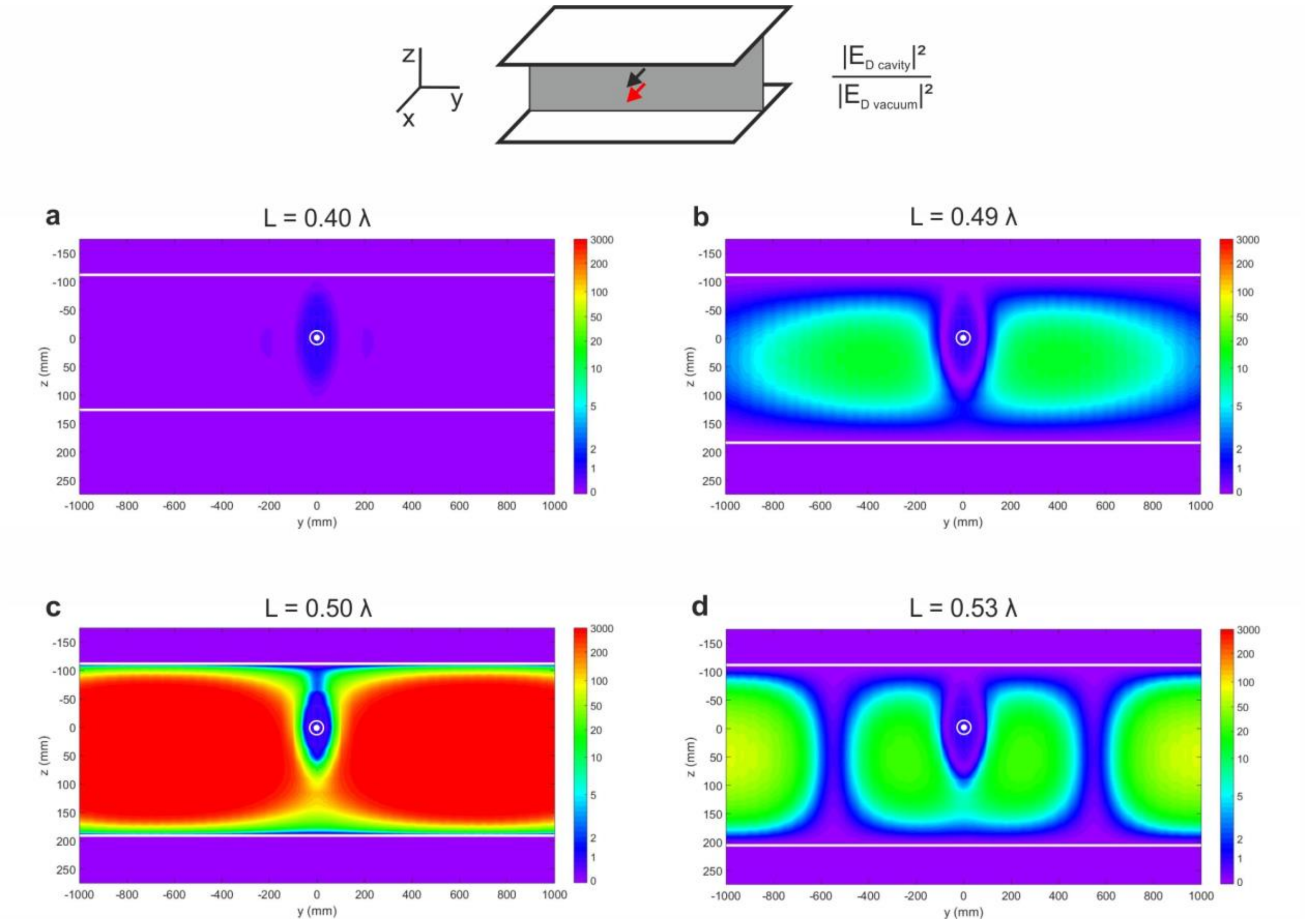

**Figure S3.** Large scale evolution of the dipole-dipole energy transfer enhancement inside the cavity. The simulations focus on the same cases as in Fig. 3 displayed here on a larger spatial range. The emission frequency is 500 MHz corresponding to a wavelength of 60 cm. The white dot indicates the donor dipole position and the white lines show the mirrors positions. All maps share the same colorscale.

## 5. FRET efficiency enhancement inside a photonic cavity

Here we focus on the role of the photonic cavity on the FRET efficiency $E_{FRET}$ defined as the probability for the donor to transfer its energy to the acceptor over all other possible decay paths:

$$E_{FRET} = \frac{\Gamma_{DA}}{\Gamma_{DA} + \Gamma_D} \tag{S7}$$

Where $\Gamma_{DA}$ is the FRET rate constant and $\Gamma_D$ is the donor's total decay rate constant in absence of the acceptor. It follows from our definition of the energy transfer enhancement that $\Gamma_{DA} = F_{ET}\ \Gamma_{DA}^0$ where the superscript $^0$ refers to the free space reference without the cavity. Likewise we can use the definition of the Purcell factor $F_P$ so that $\Gamma_D = F_P\ \Gamma_D^0$ and we can rewrite Eq. (S7) as

$$E_{FRET} = \frac{F_{ET}\ \Gamma_{DA}^0}{F_{ET}\ \Gamma_{DA}^0\ + F_P\ \Gamma_D^0} \tag{S8}$$

The definition of the Förster's radius $R_0$ relates the decay rate constants for the free space reference:

$$\frac{\Gamma_{DA}^0}{\Gamma_D^0} = \left(\frac{R_0}{R_{DA}}\right)^6 = \frac{E_{FRET}^0}{1 - E_{FRET}^0} \tag{S9}$$

Finally inserting Eq. (S9) into (S8) we get the expression of the FRET efficiency inside the cavity as a function of the energy transfer rate enhancement, Purcell factor and initial FRET efficiency:

$$E_{FRET} = \frac{E_{FRET}^0\ F_{ET}}{E_{FRET}^0\ F_{ET}\ + (1 - E_{FRET}^0) F_P} \tag{S10}$$

This equation has several practical consequences on the FRET enhancement. To illustrate the discussion, Fig. S5 shows the simulation results for three different donor-acceptor separations corresponding to high, medium and low FRET efficiency for the free-space reference. Interestingly, when the cavity length is set well below the resonance $L < \lambda/2$ then the LDOS is quenched and the Purcell factor tends to zero. However, we have seen that in these conditions the energy transfer remains mostly unaffected and $F_{ET} \approx 1$. This means that the FRET efficiency $E_{FRET}$ inside the cavity becomes close to 100% independently of the donor-acceptor separation (Fig. S5a). As the spontaneous emission is prohibited, the donor has no other choice than to give its energy to the acceptor. Such situation is highly beneficial to the cases where the initial FRET efficiency is low so that the apparent gain in FRET efficiency can be as high as $1/E_{FRET}^0$ (Fig. S5b). In return, when the cavity length is above the resonance $L > \lambda/2$ then the LDOS is enhanced the Purcell factor is greater than one. This tends to lower the FRET efficiency inside the cavity as the Purcell factor can be greater than the FRET rate enhancement.

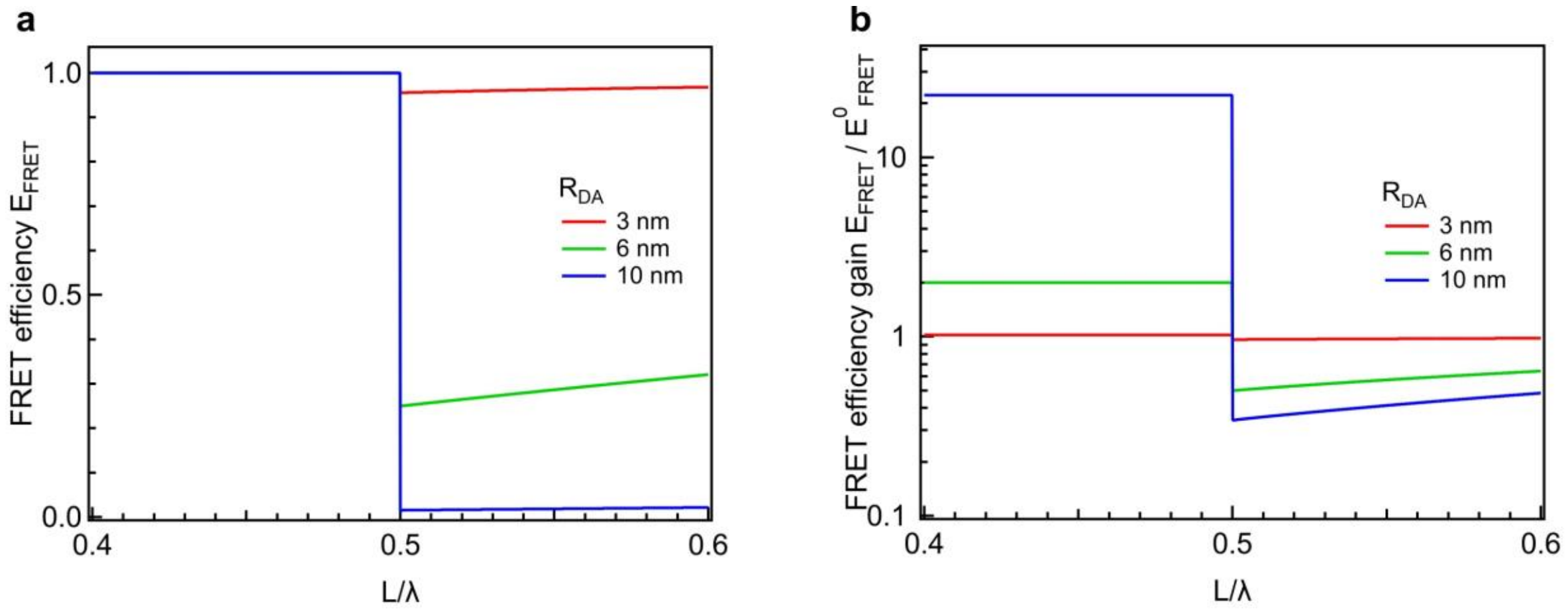

**Figure S4.** (a) Simulations of the FRET efficiency $E_{FRET}$ inside the cavity for different donor-acceptor separations $R_{DA}$. The Förster radius $R_0$ was taken equal to 6 nm so that the reference FRET efficiencies for the 3, 6 and 10 nm separations are respectively 98%, 50% and 4.5%. (b) Enhancement factor for the FRET efficiency inside the cavity derived from the data in (a).